\documentclass[11pt]{article}
\pdfoutput=1
\usepackage{tensor, physics, caption, subcaption} 
\usepackage[dvipsnames]{xcolor}
\usepackage{amsfonts}
\usepackage{amssymb}
\usepackage{mathrsfs}
\usepackage{graphicx}
\usepackage{amsmath,amsthm,amssymb,amscd}
\usepackage[all]{xy}
\usepackage{multirow}
\usepackage{color}
\usepackage{float}
\usepackage{mathtools,slashed,tensor}
\usepackage{bbold}
\usepackage{bbm}
\usepackage{slashed}
\usepackage{cancel}
\usepackage{cite, comment}
\usepackage{mdframed}
\usepackage{jheppub}

\def\be{\begin{equation}}
\def\ee{\end{equation}}
\def\bea{\begin{eqnarray}}
\def\eea{\end{eqnarray}}

\preprint{LITP-25-12}

\title{One-loop Corrected Holographic Shear Viscosity to Entropy Density Ratio at Low Temperatures}

\author{Leopoldo A.~Pando Zayas}

\author{ and Jingchao Zhang}

\emailAdd{ lpandoz@umich.edu, jingchaz@umich.edu }

\affiliation{Leinweber Institute for Theoretical Physics, 
University of Michigan, Ann Arbor, MI 48109, USA}

\abstract{Near-extremal black holes contain infrared-enhanced quantum fluctuations in their near-horizon near-AdS$_2$ throat, governed in part by Schwarzian modes. We study the effect of these fluctuations on the zero-frequency stress-tensor response of a near-extremal asymptotically AdS$_4$ Reissner--Nordstr\"om black brane whose transverse directions are regulated by a finite toroidal quotient. Working directly in four dimensions, we compute the leading correction in the weakly coupled Schwarzian regime $T_q\ll T\ll r_0/L^2$. At one loop, the shear perturbation couples to the $n=2$ would-be zero mode and produces a correction proportional to $T_q/T$. Together with the logarithmic correction to the entropy, this yields a finite-volume correction to $\eta/s$. Our calculation provides a direct four-dimensional benchmark, complementary to exact two-dimensional treatments, while keeping the asymptotic AdS$_4$ source explicit.}

\begin{document}

\maketitle

\section{Introduction}

The AdS/CFT correspondence is a powerful framework to understand strongly coupled theories with emphasis on patterns of symmetry and symmetry breaking \cite{Maldacena:1997re}. One of its most successful applications relates to its ability to capture the principles of fluid dynamics; the AdS/CFT correspondence naturally accommodates the near-equilibrium behavior of generic matter in the long wavelength, late time approximation   \cite{Policastro:2002se,Rangamani:2009xk,Hubeny:2011hd}. A seminal result that propelled this connection was the universality of the shear viscosity to entropy density ratio in holographic theories, $\frac{\eta}{s} = \frac{1}{4\pi}$,  obtained in \cite{Kovtun:2004de} (see also the comprehensive review   \cite{Cremonini:2011iq}).

Recently, our understanding of previously puzzling aspects of near-extremal, almost zero temperature black holes has improved considerably. In particular, the effective theory governing the pattern of symmetry breaking in the near-horizon region of extremal black holes has been elucidated. This effective theory generically contains the Schwarzian and has led to a better understanding of the dynamics in the throat region as recently reviewed in \cite{Mertens:2022irh,Turiaci:2023wrh}.  The systematic treatment of such zero modes has been shown to affect the thermodynamics of near-extremal black holes \cite{Iliesiu:2020qvm,Heydeman:2020hhw,Boruch:2022tno,Iliesiu:2022onk}, including also in rotating black holes where a dimensional reduction to an effective two-dimensional theory is not available \cite{Kapec:2023ruw,Rakic:2023vhv,Maulik:2024dwq}.

In this manuscript we study how the gravitational path integral over these modes modifies stress-tensor correlation functions in the AdS/CFT correspondence. At the classical level, there is no obstruction to imposing the ultraviolet boundary condition at the asymptotic AdS$_4$ boundary while solving the fluctuation equation by matching an inner AdS$_2$ solution to an outer AdS$_4$ solution. Indeed, the standard matching formula expresses the full low-frequency retarded Green's function in terms of the infrared AdS$_2$ Green's function and ultraviolet matching coefficients \cite{Faulkner:2009wj}, and the two-dimensional analyses of \cite{Cremonini:2025yqe,Gouteraux:2025exs,Kanargias:2025vul} implement precisely this construction.

The distinction relevant for the present work arises at the quantum level. Inserting an exact quantum-corrected infrared correlator into a matching relation derived in the classical bulk theory assumes that the factorization between the throat dynamics and the ultraviolet matching data continues to hold in the required form. This is a natural and productive assumption, but it is additional quantum input. Our complementary strategy keeps the asymptotic AdS$_4$ source explicit and evaluates its coupling to the lifted zero modes directly in the four-dimensional quadratic action. This approach is also adaptable to backgrounds for which a consistent two-dimensional reduction is unavailable. The application developed below is the leading finite-volume correction to the zero-frequency shear response and to $\eta/s$ in the perturbative regime $T\gg T_q$.

\section{One-loop correction to the Euclidean path integral: a brief review}
\label{Sec: One-loop partition function}
Let us first review the one-loop contributions to the entropy of near-extremal black holes and their  effects on the corrected thermodynamics. The gravitational path integral is usually considered in the approximation where it is dominated by  the saddle point approximation given simply by the on-shell action \cite{Gibbons:1976ue}. To include one-loop corrections, we consider gravitational fluctuations to quadratic order:  
\begin{equation}
\label{Eq:one-loop added}
    Z=e^{-S_{on-shell}}\int\mathcal{D}h\, e^{-\, h\,\Delta_L\,h},
\end{equation}
where the operator $\Delta_L$ is the Lichnerowicz operator acting on small metric fluctuations, $h$. In generic circumstances, the contribution of the fluctuations is small and can be neglected. For near-extremal black holes, however, as the temperature goes to zero we find that for some modes $h=h_{zero}$ their  eigenvalues vanish. Essentially, the contribution of these zero modes to the partition function blows up, signaling an IR divergence. Therefore, at low enough temperatures, the one-loop correction is no longer negligible and competes with the tree-level on-shell action. 

To better understand this IR divergence, it is convenient to introduce a small, non-vanishing temperature via a deformation of the gravitational background metric of the form
\begin{equation}
    \bar{g}=g^{(0)}+T  g^{(1)} +\mathcal{O}(T^2),
\end{equation}
where $g^{(0)}$ is the extremal configuration, while $g^{(1)}$ is the deformation that drives the black hole away from extremality. It is $g^{(1)}$ that regulates the zero modes $h_{zero}$ and allows us to understand the physics of the system as one takes the temperature to zero.

\subsection{Near-extremal Reissner--Nordstr\"om-AdS$_4$ black brane and the scale $T_q$}
In this section, we present some details for the  Reissner-Nordstr\"om-AdS black brane in four dimensions,  following part of the established literature \cite{Iliesiu:2020qvm,Heydeman:2020hhw,Boruch:2022tno,   Iliesiu:2022onk,Kapec:2023ruw,Rakic:2023vhv,Maulik:2024dwq,Blacker:2025zca,Maulik:2025phe,Arnaudo:2025btb,Arnaudo:2024rhv,Arnaudo:2024bbd}.


Consider the Einstein-Maxwell action in 3+1 dimensions with a negative cosmological constant $\Lambda=-3/L^2$,
\begin{equation}
    I=\frac{1}{2 \kappa_4^2} \int d^4 x \sqrt{-g}\left(R-2 \Lambda-L^2 F_{\mu \nu} F^{\mu \nu}\right) .
\end{equation}
The Reissner-Nordstr\"om AdS$_4$ black brane is given by the following metric and gauge field
\begin{equation}
\label{eq:RN metric}
    \begin{aligned}
d s^2 & =g_{\mu \nu} d x^\mu d x^\nu=\frac{r^2}{L^2}\left(-f(r) d t^2+d x^2+d y^2\right)+\frac{L^2}{r^2 f(r)} d r^2, \\
A & =\frac{q}{L^2}\left(
\frac{1}{r_+}-\frac{1}{r}\right) d t, \qquad 
f(r)=1-\frac{m}{r^3}+\frac{q^2}{r^4},
\end{aligned}
\end{equation}
We denote by $r_+$ the radius of the outer horizon. The line element in \eqref{eq:RN metric} specifies the local geometry of the translationally invariant directions. Throughout the calculation we regulate these directions by the rectangular toroidal quotient
\begin{equation}
\label{eq:torus identifications}
    x\sim x+L_x,\qquad y\sim y+L_y,\qquad V\equiv L_xL_y.
\end{equation}
The strict planar geometry is recovered by taking $V\to\infty$. The horizon area is $\mathcal{A}_H=(r_+^2/L^2)V$, so $V$ is also the spatial volume with respect to the conformal boundary metric. With the normalization of the action used above, the parameters $m$ and $q$ determine the energy density $\varepsilon$ and electric charge density $\rho$ according to \cite{Edalati:2009bi}
\begin{equation}
\label{eq:physical mass charge}
    \varepsilon=\frac{m}{\kappa_4^2L^4},\qquad
    \rho=\frac{2q}{\kappa_4^2L^2}.
\end{equation}
For the finite torus, the total mass and charge are therefore
\begin{equation}
    \mathcal{M}=V\varepsilon=\frac{Vm}{\kappa_4^2L^4},\qquad
    \mathcal{Q}=V\rho=\frac{2Vq}{\kappa_4^2L^2}.
\end{equation}
For the noncompact planar brane, $\mathcal{M}$ and $\mathcal{Q}$ diverge and the corresponding densities are the physical finite quantities.

The Hawking temperature and chemical potential are
\begin{equation}
\label{eq:T and mu function}
    T=\frac{r_+^2}{4\pi L^2}f'(r_+),\qquad
    \mu=\frac{q}{L^2r_+}.
\end{equation}
The horizon condition gives
\begin{equation}
\label{eq:horizon mass relation}
    m=r_+^3+\frac{q^2}{r_+}.
\end{equation}
Together with the entropy density $s=2\pi r_+^2/(\kappa_4^2L^2)$, these quantities obey
\begin{equation}
    d\varepsilon=T\,ds+\mu\,d\rho.
\end{equation}

Unless explicitly stated otherwise, the analytic calculation is performed at fixed $q$. At fixed transverse volume, this is precisely the canonical ensemble because \eqref{eq:physical mass charge} implies that fixing $q$ is equivalent to fixing the physical charge $\mathcal{Q}$. The choice is made for algebraic convenience and is not a new ensemble. At extremality,
\begin{equation}
    q=\sqrt{3}\,r_0^2,\qquad m_0=4r_0^3,
\end{equation}
where $r_0$ is the extremal horizon radius. We have also checked the fixed-$\mu$ deformation: after expressing the answer in terms of $r_0$, the temperature-induced eigenvalue shift \eqref{Eq: Lifted eigenvalue by T} is unchanged. The source-induced shift is evaluated on the exact extremal geometry and is likewise independent of the thermodynamic ensemble. Thus the coefficient of the logarithmic determinant and the explicit $1/T$ correction to the shear response are ensemble-independent at this order. The tree-level relation among $T$, $q$, $\mu$, and $r_+$, as well as the thermodynamic convention used to define the scale inside the logarithm, must be changed in the usual way when the ensemble is changed. We now expand $m$ and $r_+$ at fixed $q$ in powers of $T$:
\begin{equation}
\label{Eq: m rp expantion}
    m=m_0+\frac{3}{2}L^4\pi^2r_0\,T^2+\mathcal{O}(T^3), \qquad 
    r_+=r_0+\frac{L^2\pi}{3}T+\frac{L^4\pi^2}{6r_0}T^2+\mathcal{O}(T^3).
\end{equation}
The semi-classical entropy density of the black brane is given by
\begin{equation}
\label{Eq: Tree entropy}
    s_{tree}=\frac{2\pi}{\kappa_4^2}\left(\frac{r_0}{L}\right)^2\left(1+\frac{2\pi L^2}{3r_0}T+\mathcal{O}(T^2)\right).
\end{equation}
The linear in temperature behavior of the entropy versus the quadratic behavior of the mass was identified in \cite{Preskill:1991tb} as leading to the breakdown of thermodynamics and highlights the need to incorporate quantum corrections as the temperature is sent to zero  \cite{Iliesiu:2020qvm,Heydeman:2020hhw, Boruch:2022tno}.

To zoom into the near-horizon extremal geometry, let us consider the following coordinate transformation
\begin{equation}
\label{eq:coordinate transformation}
    r\to r_++\frac{L^2\pi}{3}\,T\,(Y-1),\qquad t\to \frac{t}{2\pi T}.
\end{equation}
For the exterior region $r\geq r_+$, the new radial coordinate has the range
\begin{equation}
    1\leq Y<\infty,
\end{equation}
with the horizon at $Y=1$ and the asymptotic end of the AdS$_2$ throat at $Y\to\infty$. We then rotate to Euclidean time
\begin{equation}
    t\to-i\tau.
\end{equation}
By taking the $T\to 0$ limit, we obtain the $AdS_2\times \mathbb{T}^2$ near-horizon extremal geometry
\begin{equation}
\label{eq:Near-horizon extremal RN-AdS}
    g^{(0)}_{\mu\nu}dx^\mu dx^\nu=\frac{L^2}{6}\left(\left(Y^2-1\right)d\tau^2+\frac{dY^2}{Y^2-1}\right)+\frac{r_0^2}{L^2}(dx^2+dy^2),\quad A^{(0)}_\mu dx^\mu=\frac{-iY}{2\sqrt{3}}d\tau.
\end{equation}
The linear in temperature terms in the background that serve to lift the zero modes are 
\begin{equation}
\label{Eq: linear T metric}
  g^{\left(1\right)}_{\mu \nu} dx^\mu dx^\nu=T\frac{2L^4\pi}{27r_0}\Bigl[-(Y-1)^2(Y+2)d\tau^2+\frac{(Y+2)dY^2}{(Y+1)^2}+\frac{9\,r_0^2\,Y}{L^4}(dx^2+dy^2)\Bigr],
\end{equation}
\begin{equation}
    A^{(1)}_\mu dx^\mu=T\frac{iL^2\pi(2Y^2-3)}{12\sqrt{3}r_0}d\tau.
\end{equation}
The Lichnerowicz operator is schematically given by
\begin{eqnarray}
    h_{\alpha\beta}^*\Delta^{\alpha \beta, \mu \nu}_{L}h_{\mu\nu}=-\frac{1}{2\kappa_4^2}h^*_{\alpha\beta}(\Delta^{\alpha \beta, \mu \nu}_{EH}-2\Lambda\Delta^{\alpha \beta, \mu \nu}_{1}-2L^2\Delta^{\alpha \beta, \mu \nu}_{F})h_{\mu\nu},
\label{eq:Linch operator}
\end{eqnarray}
where $\Delta^{\alpha \beta, \mu \nu}_{EH},\Delta^{\alpha \beta, \mu \nu}_{1},\Delta^{\alpha \beta, \mu \nu}_{F}$ are the contribution from the Einstein-Hilbert, cosmological constant and the Maxwell action, respectively. The precise forms can be found in Appendix \ref{appendix:Lich operator}. We absorb Newton's constant $\tfrac{1}{2\kappa_4^2}$ into $\Delta_L$ for convenience.
At zero temperature, i.e., when $g=g^{(0)}, A=A^{(0)}$, one can verify that the above operator admits the following zero modes:
\begin{equation}
h_{\mu\nu}^{(n)}dx^\mu dx^\nu=C_n e^{i n \tau} \left(\frac{-1 + Y}{1 + Y}\right)^{\frac{{n}}{2}} (-d\tau^2+2i\frac{d\tau dY}{Y^2-1}+\frac{dY^2}{(Y^2-1)^2})
\label{eq:zero modes}
\end{equation}
for $n\geq2$, where $C_n$ are normalization constants. The above tensor zero modes are generated by the following large diffeomorphisms:
\begin{equation}
        \xi^{(n)\mu}\frac{\partial}{\partial x^\mu}=e^{in\tau}\left(f_2(y)\frac{\partial}{\partial\tau}+f_1(y)\frac{\partial}{\partial Y}\right),
\end{equation}
where
\begin{equation}
    f_1=\frac{3}{L^2(n^2-1)}(n+Y)\left(\frac{Y-1}{Y+1}\right)^{\frac{n}{2}},\quad
    f_2=\frac{if_1'(Y)}{n}.
\end{equation}
These vector fields are non-normalizable, and change the asymptotics of the background by $\mathcal{L}_{\xi}g^{(0)}_{\mu\nu}$. The explicit expression of such would-be zero modes in the four-dimensional case and their connection with the Schwarzian modes was discussed in \cite{Iliesiu:2022onk}  by direct comparison with the original interpretation in \cite{Maldacena:2016upp,Jensen:2016pah}. In the case of rotating black holes the situation is conceptually clearer because a dimensional reduction is not available and the JT physics appears directly in four dimensions \cite{Kapec:2023ruw,Rakic:2023vhv,Maulik:2024dwq}.

We normalize the zero modes by
\begin{equation}
    \frac{1}{2\kappa_4^2}\int d^4x\sqrt{-g}\,h^*_{\alpha\beta}h^{\alpha\beta}=l^2,
\end{equation}
where $l$ is a scaling parameter with dimension of length,  $l$ does not appear in physical quantities. The normalization gives
\begin{equation}
    C_n^2=\frac{l^2L^4n(n^2-1)\kappa^2_4}{6\pi r_0^2 V},
\end{equation}
where $V=L_xL_y$ is the finite transverse volume introduced in \eqref{eq:torus identifications}.

By turning on a small temperature, the ``zero modes'' are lifted and acquire nonzero eigenvalues. Following perturbation theory, we substitute $\bar{g}=g^{(0)}+ g^{(1)}$ into the Lichnerowicz operator \eqref{eq:Linch operator} and apply it to the zero modes \eqref{eq:zero modes}. The lifted eigenvalues are
\be
\label{Eq: integral}
\delta\Lambda_n=\int dx^4 \sqrt{g^{(0)}}h^{(n)*}_{\alpha\beta}\delta\Delta_L^{\alpha\beta,\mu\nu} h^{(n)}_{\mu\nu}.
\ee
The integration leads to a simple  result
\be
\label{Eq: Lifted eigenvalue by T}
\delta\Lambda_n=\frac{2\pi\, l^2}{r_0} \,n\,T,\quad   n\geq2.
\ee
More details for  this calculation can be found in Appendix \ref{appendix:Lich operator}. The contribution of the extremal zero modes to the low-temperature partition function is evaluated via zeta-function regularization leading to 
\be
\delta \log Z=\log(\prod_{n\geq2}\frac{\pi}{\delta\Lambda_n})=\frac{3}{2}\log (\frac{T}{T_q})+{\cal O}(1).
\ee
The coefficient of this logarithmic correction  $(3/2)$ is quite robust. The ${\cal O}(1)$ term is ambiguous since it gets contributions from non-zero modes and depends on the precise normalization of the partition function. We, therefore, find  $T_q$ by physical considerations. Namely, $T_q$ is the temperature scale under which the semi-classical thermodynamics breaks down as predicted in \cite{Preskill:1991tb}. This happens when the semi-classical energy of the black hole above extremality is comparable to or even smaller than the average energy of a Hawking quantum. This temperature can be found by the inverse of the ${\cal O}(T)$ coefficient of the classical entropy \eqref{Eq: Tree entropy} as
\be
\label{Eq: tq}
T_q=\frac{4\pi^2}{s^{(1)} V}=\frac{4\pi^2 T}{C_q(T)}=\frac{3\kappa_4^2}{r_0 V},
\ee
where the $4\pi^2$ is a convention, $C_q(T)=s^{(1)}V T$ is the semi-classical heat capacity with constant $q$ at low temperature. The classical low-temperature expansion becomes strongly affected by the Schwarzian sector around the scale $T_q$. In the perturbative regime $T\gg T_q$, the corresponding one-loop correction to the thermodynamic entropy is
\begin{equation}
    \label{Eq: one-loop S}
    S_{\rm one\mbox{-}loop}=\frac{3}{2}\log\left(\frac{T}{T_q}\right),\qquad
    s_{\rm one\mbox{-}loop}=\frac{S_{\rm one\mbox{-}loop}}{V}
    =\frac{3}{2V}\log\left(\frac{T}{T_q}\right).
\end{equation}
The finite volume is essential here. Since $T_q\propto V^{-1}$, at fixed nonzero $T$ one has $s_{\rm one\mbox{-}loop}\sim (\log V)/V\to0$ as $V\to\infty$. The formula above captures the infrared-enhanced contribution of the spatially homogeneous Schwarzian sector; it is not a statement that every bulk one-loop correction vanishes in the thermodynamic limit.
\section{AdS/CFT correlation function: from tree level to one-loop formalism}
\label{Sec: Formalism}

We aim to incorporate the effects of the zero modes in the broader context of the AdS/CFT correspondence. To implement this, we turn to the main formula 
\begin{equation}
    Z_{\rm CFT}[J]=Z_{\rm bulk}[\phi_0],
\end{equation}
where $\phi_0$ is the boundary value of the bulk field dual to the operator that couples to the source $J$.

\subsection{One-loop corrected correlation functions}
Let us first consider the original AdS/CFT dictionary
\begin{equation}
\left\langle \exp\left( \int \mathrm{d}^d x\, \mathcal{O}\,\phi_{(0)} \right) \right\rangle_{\mathrm{CFT}}
= Z_{\mathrm{string}} \Bigg|_{\displaystyle \lim_{z\to0}\,\bigl(\,\phi(z,x)\, z^{\Delta - d}\bigr) = \phi_{(0)}(x)} \,.
\end{equation}
At tree level, we use the saddle point to approximate the partition function in the bulk by the semiclassical on-shell action. This is the approximation where most of the AdS/CFT correspondence is applied. 


Several works have incorporated Schwarzian dynamics into holographic correlation functions \cite{Daguerre:2023cyx,Liu:2024gxr,Liu:2024qnh,Castro:2025pst,Jiang:2025cyl}. In the classical matching construction, the asymptotic AdS$_4$ boundary condition is imposed in the outer region and the low-frequency nonanalyticity is supplied by the AdS$_2$ inner-region Green's function \cite{Faulkner:2009wj}. The issue we wish to isolate is therefore not whether the ultraviolet boundary condition can be imposed; it can. Rather, at the quantum level one must decide how the quantum throat correlator is embedded into the full ultraviolet observable. The exact two-dimensional approaches assume that the classical matching structure can be used with the quantum-corrected infrared correlator. Our four-dimensional calculation provides an independent perturbative implementation: the bulk source is first normalized at the AdS$_4$ boundary, and its contribution to the lifted eigenvalues is then evaluated directly in the four-dimensional quadratic action.

As $T$ approaches $T_q$, the saddle-point approximation becomes strongly coupled. We restrict instead to $T\gg T_q$, where \eqref{Eq:one-loop added} gives the leading quantum correction. Whether a boundary correlator receives a contribution from the Schwarzian sector is determined by whether the corresponding bulk source $\phi$ changes the quadratic form $h_{\rm zero}\,\Delta_Lh_{\rm zero}$.

We first add the one-loop correction from the zero modes to the AdS/CFT dictionary
\begin{equation}
    \left\langle \exp\left( \int \mathrm{d}^d x\, \mathcal{T}\,\delta g_{(0)} \right) \right\rangle_{\mathrm{CFT}}
=
e^{-S_{\mathrm{on-shell}}}\int \mathcal{D}h\;e^{-\,h_{\rm zero}\,\Delta_{L}\,h_{\rm zero}}
\Bigg|_{\displaystyle \lim_{z\to0}\,\bigl(\,\delta g(z,x)\, z^{\Delta - d}\bigr)\,=\,\delta g_{(0)}(x)} \,.
\end{equation}
The metric field is given by 
\begin{equation}
\label{Eq: metric perturbation}
    g=\bar{g}+\delta g=g^{(0)}+ g^{(1)}+\delta g.
\end{equation}
The notation indicates that $g^{(0)}$ is the metric at zero temperature, $g^{(1)}$ is the first order in temperature metric when one turns on a small temperature and $\delta g$ denotes the bulk field dual to the field theory operator $\mathcal{T}$. When we study the correlation function in AdS/CFT, it is the combination $\bar{g}=g^{(0)}+ g^{(1)}$ that serves as the background metric.

We already established that  $g^{(1)}$  contributes to $h_{\rm zero}\,\Delta_{L}\,h_{\rm zero}$. The key  question now is whether $\delta g$ contributes to $h_{\rm zero}\,\Delta_{L}\,h_{\rm zero}$ and how this contribution depends on the boundary data $\delta g_{(0)}$.  For example, the two-point correlation function is 
\begin{equation}
\label{Eq: Two-Point function}
    \langle\mathcal{T}(x_1)\mathcal{T}(x_2)\rangle=\frac{1}{Z}\frac{\delta Z}{\delta \,(\delta g_{(0)}(x_1))\delta \,(\delta g_{(0)}(x_2))}
\Bigg|_{\displaystyle \lim_{z\to0}\,\bigl(\,\delta g(z,x)\, z^{\Delta - d}\bigr)\,=\,\delta g_{(0)}(x)} \,.
\end{equation}
where we incorporate one-loop corrections in $Z$:
\begin{equation}
    Z=e^{-S_{\mathrm{on-shell}}}\int \mathcal{D}h\;e^{-\,h_{\rm zero}\,\Delta_{L}\,h_{\rm zero}}.
\end{equation} 
Let us focus on the one-loop part at small but non-vanishing temperature,
\begin{equation}
    \int \mathcal{D}h\;e^{-\,h_{zero}\,\Delta_{L}\,h_{zero}}\sim (\det \Delta_L)^{-1}=(\prod_n\delta\Lambda_n)^{-1},
\end{equation}
where $\delta\Lambda_n$ is the lifted eigenvalue of the $n$th zero mode. When the metric is given by \eqref{Eq: metric perturbation}, $\delta\Lambda_n$ comes from two parts
\begin{equation}
\label{Eq: whole lifted eigenvalue}
    \delta \Lambda_n=\delta\Lambda_n[g^{(1)}]+\delta\Lambda_n[\delta g].
\end{equation}
The contribution $\delta\Lambda_n[\delta g]$ determines the one-loop correction to the correlation function. As both $\delta g$ and $g^{(1)}$ are small perturbations with respect to $g^{(0)}$, we can calculate $\delta\Lambda_n[\delta g]$ and $\delta\Lambda_n[g^{(1)}]$ separately and simply add them together. For near-extremal Reissner-Nordstr\"om-AdS black brane, $\delta\Lambda_n[g^{(1)}]$ is given by \eqref{Eq: Lifted eigenvalue by T}. We will calculate the crucial new ingredient, $\delta\Lambda_n[\delta g]$,  in Section \ref{Sec: One-loop eta}. 
\section{Shear viscosity: non-zero temperature and one-loop correction}
In this section, we apply the formalism sketched in Section \ref{Sec: Formalism} to the explicit example of a Reissner-Nordstr\"om-AdS black brane. Previous computations that considered the shear viscosity to entropy ratio for extremal backgrounds, took $T\to 0$ as a first step and did not, consequently, include the low-energy Schwarzian-like modes \cite{Edalati:2009bi, Edalati:2010hk, Edalati:2010pn}. Here, we keep a small finite temperature throughout the computation.

\subsection{Small-temperature expansion at tree level}
Before turning to the one-loop correction, we first discuss the tree-level retarded Green's function. This subsection extends the extremal calculation of \cite{Edalati:2009bi} to linear order in temperature. Similar to \cite{Edalati:2009bi}, in this paper we consider the limit of $\vec{k}=0$ and $\omega\to0$.

To study the shear viscosity of the boundary theory, we start with the following metric fluctuations
\begin{equation}
\label{Eq: phi to gxy}
    \delta g_{\mu\nu}dx^\mu dx^\nu=2\,e^{-i\omega t}\; h^x{}_{y}(r)g_{xx}\,dxdy\equiv2\, e^{-i\omega t}\phi(r)\frac{r^2}{L^2}dxdy
\end{equation}
on the background of \eqref{eq:RN metric}, where we have defined $\phi(r)\equiv h^x{}_y(r)$ for the convenience of notation and also to distinguish with the variable $h$ of the zero modes \eqref{eq:zero modes}. 

Direct algebra shows the linear Einstein equation for $\phi(r)$:
\begin{equation}
\label{Eq: Full ODE}
\phi''(r)
+ \frac{m - 4 r^{3}}{\,m r - r^{4} - 3 r_{0}^{4}\,}\,\phi'(r)
+ \frac{L^{4} r^{4}\omega^{2}}{\big(m r - r^{4} - 3 r_{0}^{4}\big)^{2}}\,\phi(r)
= 0,
\end{equation}
where $m$ depends on temperature $T$. We will solve this ODE up to order $T$ in two regions:
\begin{itemize}
    \item Inner region: the near-horizon region described by \eqref{eq:Near-horizon extremal RN-AdS} and \eqref{Eq: linear T metric}. In this region, $\frac{r-r_+}{r_+}\ll1$.
    \item Outer region: where $\tfrac{(r-r_+)^2}{L^2r_+\omega}\gg1$. In this region we can neglect the $\phi(r)$ term in \eqref{Eq: Full ODE}. 
\end{itemize}
 Let us consider $\omega\ll\tfrac{r_+}{L^2}$. These two regions have overlap at $L^2 r_+\omega\ll (r-r_+)^2\ll r_+^2$.  We will solve \eqref{Eq: Full ODE} in two regions separately then connect them in the overlap region.

\paragraph{Inner region:} Applying the coordinate transformation \eqref{eq:coordinate transformation} to \eqref{Eq: Full ODE} gives the inner-region equation. There are two independent small scales, $T$ and $\omega$. The observable considered here is the Kubo coefficient at a fixed, nonzero temperature,
\begin{equation}
    \eta=-\lim_{\omega\to0}\frac{1}{\omega}\operatorname{Im}G^R_{xy,xy}(\omega,\vec{k}=0).
\end{equation}
Accordingly, the order of limits relevant for this paper is to keep $T>0$ while taking the honest zero-frequency limit. Defining
\begin{equation}
    \Omega\equiv\frac{\omega}{2\pi T},
\end{equation}
this prescription implies $\Omega\to0$. The choice is not a restriction intrinsic to the determinant method: limits with fixed or large $\omega/T$ probe finite-frequency response functions rather than the Kubo viscosity. Keeping $T$ nonzero also regulates the Schwarzian zero modes, for which perturbation theory around the strict extremal saddle is singular. Up to linear order in $T$, the inner equation is
\begin{equation}
\label{Eq: inner ODE}
\mathcal{E}^{(0)}+T\mathcal{E}^{(1)}=0
\end{equation}
with
\begin{equation}
    \mathcal{E}^{(0)}(\phi_{\mathcal{I}})=(Y^{2}-1)\,\phi_{\mathcal{I}}''(Y)
+ 2Y\,\phi_{\mathcal{I}}'(Y)
+ \frac{\Omega^{2}}{Y^{2}-1}\,\phi_{\mathcal{I}}(Y)
\end{equation}
\begin{equation}
    \mathcal{E}^{(1)}(\phi_{\mathcal{I}})=\frac{2\pi L^{2}}{9\,r_{0}}
\!\left[
(-4+3Y+Y^{3})\,\phi_{\mathcal{I}}''(Y)
+ 3(1+Y^{2})\,\phi_{\mathcal{I}}'(Y)
+ \frac{(-4+5Y+5Y^{2})\,\Omega^{2}}{(Y-1)(1+Y)^{2}}\,\phi_{\mathcal{I}}(Y)
\right]\!.
\end{equation}
where $\phi_{\mathcal{I}}$ denotes the solution in the inner region. We first solve the $\mathcal{O}(1)$ equation,
\begin{equation}
    \mathcal{E}^{(0)}(\phi_{\mathcal{I}}^{(0)})=0.
\end{equation}
This gives
\begin{equation}
    \phi_{\mathcal{I}}^{(0)}(Y)=c_1\left( \frac{-1 + Y}{1 + Y} \right)^{-\frac{i \Omega}{2}}+c_2\left( \frac{-1 + Y}{1 + Y} \right)^{+\frac{i \Omega}{2}},
\end{equation}
where $c_1,c_2$ are constants.
Imposing in-going boundary condition at the horizon $Y=1$ selects $c_2=0$:
\begin{equation}
\label{Eq: Inner hxy}
    \phi_{\mathcal{I}}^{(0)}(Y)=c_1\,\left( \frac{-1 + Y}{1 + Y} \right)^{-\frac{i \Omega}{2}}.
\end{equation}
Then we solve the $\mathcal{O}(T)$ part. Substituting $\phi=\phi^{(0)}+T\phi^{(1)}$ into \eqref{Eq: inner ODE}, we get the equation for $\phi^{(1)}$ 
\begin{equation}
    \mathcal{E}^{(0)}(\phi_{\mathcal{I}}^{(1)})+\mathcal{E}^{(1)}(\phi_{\mathcal{I}}^{(0)})=0,
\end{equation}
which is an inhomogeneous equation. The homogeneous solution is identical to \eqref{Eq: Inner hxy} after imposing the ingoing boundary condition, while a particular solution can be obtained by variation of constants.
\begin{equation}
    \phi_{\mathcal{I}}(Y)=(c_1+T\tilde{c}_1)\;\,\left( \frac{-1 + Y}{1 + Y} \right)^{-\frac{i \Omega}{2}}+T \varphi(Y,c_1),
\end{equation}
where $\varphi(Y,c_1)$ is the particular solution which depends on $c_1$. This particular solution has a complicated expression. However, the role of the inner-region solution is to impose the in-going boundary condition and to connect with the outer-region solution. Therefore, we only care about the expression of $\varphi$ in the connecting region where $Y\to\infty$. Expanding to first order in $\frac{1}{Y}$ and taking $\Omega\to0$ shows
\begin{equation}
    \varphi(Y,c_1)=\frac{2c_1L^2\pi}{9r_0}\left(-i\Omega\left(1+\log{2}\right)+i\Omega\log{Y}+\frac{3i\Omega}{Y}+\dots\right),
\end{equation}
where the dots are terms of order  $\mathcal{O}(\frac{1}{Y^2})$ or $\mathcal{O}(\Omega^2)$. Therefore, in the matching region, $\phi_{\mathcal{I}}$ goes like
\begin{equation}
\label{Eq: phi_in match in Y}
\begin{aligned}
        \phi_{\mathcal{I}}=&\frac{2\pi\, i\,c_1L^2T\Omega}{9r_0}\log{Y}+\left(c_1+\tilde{c}_1T-\frac{2\pi i\,c_1L^2T\Omega(1+\log{2})}{9r_0}\right)\\
        &+\left(i\Omega(c_1+\tilde{c}_1T)+\frac{2\pi i \,c_1L^2T\Omega}{3r_0}\right)\frac{1}{Y}+\mathcal{O}(\frac{1}{Y^2}),
\end{aligned}
\end{equation}
where we have also eliminated terms of higher order in $\Omega$.

\paragraph{Outer region:} In the outer region, $\tfrac{L^4r_+^2\omega^2}{(r-r_+)^4}\ll1$. This allows us to neglect the $\phi(r)$ term at the $\omega\to0$ limit. Following the convention of \cite{Edalati:2009bi}, we use $u=r/r_0$ as the radial coordinate in the outer region. One can verify that the ODE for $\phi$ in the outer region receives correction from $\mathcal{O}(T^2)$. The ODE up to $\mathcal{O}(T)$ is the same as that at $T=0$.
\begin{equation}
    (3-4u+u^4)\phi_{\mathcal{O}}''(u)+4(u^3-1)\phi_{\mathcal{O}}'(u)=0,
\end{equation}
which has a solution of the general form 
\begin{equation}
\phi_{\mathcal{O}}(u) = a^{(0)}_{\mathcal{O}} + \frac{1}{36} b^{(0)}_{\mathcal{O}} \left[ -\frac{6}{u - 1} - 4 \log(u - 1) + \sqrt{2} \tan^{-1} \left( \frac{u + 1}{\sqrt{2}} \right) + 2 \log(u^2 + 2u + 3) \right].
\end{equation}
Near the boundary $u\to+\infty$, 
\begin{equation}
\label{Eq: phi Out AdS boundary}
\phi_{\mathcal{O}}(u) \big|_{u \to \infty} = \left( a^{(0)}_{\mathcal{O}} + \frac{\pi}{36 \sqrt{2}} b^{(0)}_{\mathcal{O}} \right) - \frac{b^{(0)}_{\mathcal{O}}}{3 u^3} + \cdots,
\end{equation}
while near the matching region, $u\to1$, it becomes
\begin{equation}
\label{Eq: Matching Out}
\phi_{\mathcal{O}}(u) \big|_{u \to 1} = -\frac{b^{(0)}_{\mathcal{O}}}{6(u - 1)} \left\{ 1 + \cdots \right\} + \left[ a^{(0)}_{\mathcal{O}} + \frac{b^{(0)}_{\mathcal{O}}}{36} \left( \sqrt{2} \tan^{-1}(\sqrt{2}) + \log(36) \right) \right] \left\{ 1 + \cdots \right\}.
\end{equation}

\paragraph{Match solutions in the two regions}
In order to obtain the boundary data of the perturbation, let us now do the matching between the ``inner" and ``outer" regions.

Noticing $r=r_0+k_1 T\,Y$ and $u=\frac{r}{r_0}$, we can rewrite \eqref{Eq: phi_in match in Y} in terms of $u$ 
\begin{equation}
\label{Eq: Matching In}
\begin{aligned}
        \phi_{\mathcal{I}}(Y) \big|_{Y\to\infty}=&\frac{2\pi\, i\,c_1L^2T\Omega}{9r_0}\log{\frac{r_0(u-1)}{k_1T}}+\left(c_1+\tilde{c}_1T-\frac{2\pi i\,c_1L^2T\Omega(1+\log{2})}{9r_0}\right)\\
        &+\left(i\Omega(c_1+\tilde{c}_1T)+\frac{2\pi i\,c_1L^2T\Omega}{3r_0}\right)\frac{k_1T}{r_0(u-1)}+\mathcal{O}(\frac{1}{(u-1)^2}),
\end{aligned}
\end{equation}
Matching \eqref{Eq: Matching Out} and \eqref{Eq: Matching In} gives
\begin{equation}
\begin{aligned}
    a^{(0)}_{\mathcal{O}} =&\,c_1+\tilde{c}_1T+\mathcal{O}(T^2), \\
b^{(0)}_{\mathcal{O}} =& -\frac{2\pi i\, c_{1} L^{2} \Omega T}{r_{0}}
-\frac{2 \pi\left(2\pi i\, c_{1} L^{4}  \Omega
+ 3 i c_{1}' L^{2}  r_{0} \Omega \right) T^{2}}{3 r_{0}^{2}}+\mathcal{O}(T^3),
\end{aligned}
\end{equation}
where we have neglected terms of higher order in $\Omega$.
Now \eqref{Eq: phi Out AdS boundary} shows the asymptotic form of $h^x{}_y(u)$ near the AdS boundary. Reading off the normalization of $h^x{}_y(u)$ from the Einstein--Hilbert action together with the Gibbons--Hawking term and applying the real-time recipe of \cite{Son:2002sd}, we have
\begin{equation}
\label{Eq: tree gree}
    G^{R}_{xy,xy}(\omega,0)
= -\frac{1}{2 \kappa_{4}^{2}}
\left( \frac{r_{0}}{L} \right)^{2}
\left(i\omega+\frac{2\pi \,i \,\omega L^2T}{3r_0}+\mathcal{O}(T^2)\right),
\end{equation}
where we have used $\Omega=\frac{\omega}{2\pi T}$. Applying the Kubo formula
\begin{equation}
\label{Eq: eta tree}
    \eta_{tree} = - \lim_{\omega \to 0} \frac{1}{\omega} 
\, \mathrm{Im} \, G^{R}_{xy,xy}(\omega,0)=\frac{1}{2 \kappa_{4}^{2}}
\left( \frac{r_{0}}{L} \right)^{2}\left(1+\frac{2\pi L^2T}{3r_0}+\mathcal{O}(T^2)\right).
\end{equation}
Recalling the expression for the entropy at small temperature \eqref{Eq: Tree entropy}, we find, up to order $T$,
\begin{equation}
    \frac{\eta_{tree}}{s_{tree}}=\frac{\frac{1}{2 \kappa_{4}^{2}}
\left( \frac{r_{0}}{L} \right)^{2}\left(1+\frac{2\pi L^2T}{3r_0}\right)}{\frac{2\pi}{\kappa_4^2}\left(\frac{r_0}{L}\right)^2\left(1+\frac{2\pi L^2T}{3r_0}\right)}=\frac{1}{4\pi}.
\end{equation}
This explicit treatment shows that, after turning on a small temperature, the tree-level relation $\eta/s=1/(4\pi)$ continues to hold. This agrees with the universality of the ratio in two-derivative gravity \cite{Buchel:2003tz}. The expansion is valid when
\begin{equation}
    \frac{TL^2}{r_0}\ll1,
\end{equation}
and the perturbative quantum calculation below additionally requires $T/T_q\gg1$.

The standard hydrodynamic interpretation becomes delicate at very low temperature and finite volume, where the hierarchy between relaxation, wavelength, and system size can fail \cite{Emparan:2025sao}. We therefore distinguish the mathematical Kubo coefficient from a claim that a conventional hydrodynamic effective theory is valid everywhere in parameter space. The zero-momentum retarded Green's function and its $\omega\to0$ coefficient remain well-defined observables, and we use the term ``shear viscosity'' in this Kubo sense. The exact-extremal calculations of \cite{Edalati:2009bi,Edalati:2010hk,Edalati:2010pn} give $\eta/s=1/(4\pi)$ at tree level. We now determine the leading finite-volume correction from the tensor zero modes.
\subsection{One-loop contribution}
\label{Sec: One-loop eta}
As discussed in Section \ref{Sec: One-loop partition function}, the Schwarzian sector becomes strongly coupled as $T$ approaches $T_q$. We work on the perturbative side of this crossover, $T\gg T_q$, where the leading correction is the one-loop contribution computed below.

The field $\delta g$ couples to the zero modes \eqref{eq:zero modes} through the lifted eigenvalue $\delta\Lambda_n[\delta g]$ in \eqref{Eq: whole lifted eigenvalue} in the near-horizon ``throat''. Now the question is to calculate $\delta\Lambda_n[\delta g]$. We do the integral \eqref{Eq: integral} again, but substitute $g=g^{(0)}+\delta g$ instead where $\delta g$ is given by \eqref{Eq: phi to gxy} and \eqref{Eq: Inner hxy}
\begin{equation}
\label{Eq: the AdS/CFT modes}
\begin{aligned}
    \delta g_{\mu\nu}dx^\mu dx^\nu&=2\int\frac{d\omega dk_xdk_y}{(2\pi)^3}\, e^{-\Omega \tau+i k_x x+ik_y y}\phi^{(0)}_{\mathcal{I}}(Y)\frac{r_0^2}{L^2}\;dxdy\\
    &=2\int\frac{d\omega dk_xdk_y}{(2\pi)^3}\,c_1(\omega,k_x,k_y)\frac{r_0^2}{L^2}e^{-\Omega\tau+i k_x x+ik_y y}\left(\frac{Y-1}{Y+1}\right)^{\tfrac{-i\Omega}{2}}dxdy.
\end{aligned}
\end{equation}
Here we have neglected the $\mathcal{O}(T)$ solution of $\phi_{\mathcal{I}}(Y)$, as the contribution of $\phi^{(1)}_{\mathcal{I}}$ is sub-leading compared to that of $\phi^{(0)}_{\mathcal{I}}$. When conducting the integral \eqref{Eq: integral}, we need to rotate back to the Lorentzian time $t$ to pick out the right Fourier mode in $\omega$. Details are provided in Appendix \ref{appendix: Integral}. The results are
\begin{equation}
\label{Eq: Lifted Eigenvalue by fluctuation}
    \begin{aligned}
        \delta \Lambda_n[\delta  g]=&0,&n\ge3,\\
        \delta \Lambda_n[\delta  g]=&\frac{9\,i\,l^ 2\int d\omega dk_xdk_y\;\omega\,c_1(\omega,k_x,k_y)\,c_1(-\omega,-k_x,-k_y)}{(2\pi)^4L^2V},&n=2,\\
    \end{aligned}
\end{equation}
We see that $\delta g$ does couple to the $n=2$ zero mode and to no other higher zero mode. Together with \eqref{Eq: Lifted eigenvalue by T}, we find the total lifted eigenvalues are given by
\begin{equation}
    \begin{aligned}
        \delta \Lambda_n=&\frac{2\pi\,l^2 nT}{r_0},&n\ge3,\\
        \delta \Lambda_n=&\frac{4\pi\,l^2 T}{r_0}+\delta\Lambda_2[\delta g]&n=2,\\
    \end{aligned}
\end{equation}
we can now obtain the one-loop correction for $n\ge2$:
\begin{equation}
\label{Eq: Oneloop partition function}
\begin{aligned}
       \int \mathcal{D}h\;e^{-\,h_{zero}\,\Delta_{L}\,h_{zero}}&\sim \,(\prod_{n\ge2}\delta\Lambda_n)^{-1},\\
        =&\left(\frac{4\pi l^2 T}{r_0}+\delta\Lambda_2 [\delta g] \right)^{-1}\prod_{n\ge3}\delta\Lambda_n^{-1}\\
        =&\left(2\alpha T+\delta\Lambda_2 [\delta g]\right)^{-1}\sqrt{\frac{2}{\pi}}(\alpha\,T)^{\frac{5}{2}},
\end{aligned}
\end{equation}
where $\alpha=\frac{2\pi l^2}{r_0}$.
We have all the ingredients to determine how the one-loop correction influences the retarded Green's function. Applying \eqref{Eq: Two-Point function}, we have
\begin{equation}
\begin{aligned}
        G^{R}_{xy,xy}(\omega,0)=&\frac{(2\pi)^6\delta Z}{\delta \,\phi_0(-\omega,0,0)\delta \,\phi_0(\omega,0,0)}/Z\,\big|_{\phi_0=0}\\
        =&\frac{(2\pi)^6e^{-S_{\mathrm{on-shell}}}}{\delta \,\phi_0(-\omega,0,0)\delta \,\phi_0(\omega,0,0)}/e^{-S_{\mathrm{on-shell}}}\\&+\frac{(2\pi)^6\delta\int \mathcal{D}h\;e^{-\,h_{zero}\,\Delta_{L}\,h_{zero}}}{\delta \,\phi_0(-\omega,0,0)\delta \,\phi_0(\omega,0,0)}/\left(\int \mathcal{D}h\;e^{-\,h_{zero}\,\Delta_{L}\,h_{zero}}\right)\big|_{\phi_0=0}\\
        =&G^{R(tree)}_{xy,xy}(\omega,0)+G^{R(one-loop)}_{xy,xy}(\omega,0)
\end{aligned}
\end{equation}
where the boundary ``source" of $\delta g_{xy}$: $\phi_0=\displaystyle\lim_{u\to\infty}\,\phi_{\mathcal{O}}(u)\,= c_1$. The tree-level Green's function is given in \eqref{Eq: tree gree}.
Our main interest is in the second term. Substituting \eqref{Eq: Oneloop partition function}, we obtain
\begin{equation}
    G^{R(one-loop)}_{xy,xy}(\omega,0)=(2\pi)^6\frac{\delta\Bigl[\left(2\alpha T+\delta\Lambda_2 [\delta g]\right)^{-1}\sqrt{\frac{2}{\pi}}(\alpha\,T)^{\frac{5}{2}} \Bigr]/(\frac{1}{\sqrt{2\pi}}(\alpha\, T)^{\frac{3}{2}})}{\delta c_1(-\omega,0,0)\delta c_1(\omega,0,0)}\Big|_{c_1=0}=-
    \frac{9\,i \pi \,\omega\,r_0}{L^2\,T\,V},
\end{equation}
where we have used $\Omega=\frac{\omega}{2\pi T}$.
As for the entropy, the $1/T$ behavior signals the loss of perturbative control if one attempts to continue the one-loop answer all the way to $T=0$. In the regime $T\gg T_q$, the Kubo formula gives
\begin{equation}
\label{eq:eta one loop}
    \eta_{\rm one\mbox{-}loop}
    =-\lim_{\omega\to0}\frac{1}{\omega}\operatorname{Im}G^{R({\rm one\mbox{-}loop})}_{xy,xy}(\omega,0)
    =\frac{9\pi r_0}{L^2TV}
    =\frac{3\pi r_0^2T_q}{\kappa_4^2L^2T}.
\end{equation}
It is useful to compare this term with the linear-in-$T$ correction in \eqref{Eq: eta tree}. Their equality defines the bookkeeping scale
\begin{equation}
    T_q^{(\eta)}=\frac{3\sqrt{3}\,\kappa_4}{L\sqrt{V}}.
\end{equation}
For small $\kappa_4$, this scale is parametrically larger than $T_q$. This comparison does not mark a failure of the loop expansion: it only says that the leading quantum term can exceed the first thermal correction while both remain small compared with the nonzero extremal value of $\eta_{\rm tree}$.

After including the one-loop correction at small temperature, we find
\begin{equation}\label{Eq:eta-over-s}
\begin{aligned}
    \frac{\eta}{s}=&\frac{\eta_{tree}+\eta_{one-loop}}{s_{tree}+s_{one-loop}}
    =\frac{\frac{1}{2 \kappa_{4}^{2}}
\left( \frac{r_{0}}{L} \right)^{2}\left(1+\frac{2\pi L^2T}{3r_0}\right)+\frac{3\pi\,r_0^2\, T_q}{\kappa_4^2\,L^2\,T}}{\frac{2\pi}{\kappa_4^2}\left(\frac{r_0}{L}\right)^2\left(1+\frac{2\pi L^2T}{3r_0}\right)+\frac{3}{2V}\log{\frac{T}{T_q}}}\\
=& \frac{1}{4\pi}\,\,\times \,\,
\frac{1+\frac{3 \pi\,r_0^2\, T_q}{\kappa_4^2\,L^2\,T}\big[\frac{1}{2\kappa_4^2}(\frac{r_0}{L})^2\left(1+\frac{2\pi L^2T}{3r_0}\right)\big]^{-1}}{1+\frac{3}{8\pi V}\log{\frac{T}{T_q}}\big[\frac{1}{2\kappa_4^2}(\frac{r_0}{L})^2\left(1+\frac{2\pi L^2T}{3r_0}\right)\big]^{-1}}.
\end{aligned}
\end{equation}
For $T/T_q\gg1$, Eq.~\eqref{Eq:eta-over-s} approaches the classical value. Expanding consistently to first order in the one-loop contributions gives
\begin{equation}
\label{eq:eta over s expanded}
    \frac{\eta}{s}=\frac{1}{4\pi}\left[
    1+\frac{1}{1+\frac{2\pi L^2T}{3r_0}}
    \left(
       6\pi\frac{T_q}{T}
       -\frac{L^2T_q}{4\pi r_0}\log\frac{T}{T_q}
    \right)
    +\mathcal{O}({\rm two\mbox{-}loop})
    \right].
\end{equation}
This expression makes clear that the sign of the change in $\eta/s$ is determined by the competition between the $T_q/T$ correction to $\eta$ and the logarithmic correction to $s$. In the ordinary controlled near-extremal scaling, where $TL^2/r_0\ll1$ and $\log(T/T_q)$ is not parametrically as large as $r_0/(L^2T)$, the first term dominates and the ratio is shifted upward. A downward shift would require the extreme correlated scaling
\begin{equation}
\label{eq:downward}
    \log\frac{T}{T_q}>\frac{24\pi^2r_0}{L^2T}.
\end{equation}
We therefore do not interpret the perturbative formula as establishing a generic or parametrically sizable low-temperature violation of the KSS lower bound.  The robust content of the calculation is the finite-volume one-loop correction and its $T_q/T$ scaling.

\paragraph{Regime of validity and relation to the two-dimensional approach.}
The controlled hierarchy is
\begin{equation}
\label{eq:validity hierarchy}
    1\ll\frac{T}{T_q},\qquad \frac{TL^2}{r_0}\ll1.
\end{equation}
Although the Schwarzian partition function is one-loop exact \cite{Stanford:2017thb}, the source-dependent correlation function need not be. Our calculation therefore does not access the strongly coupled regime $T\ll T_q$ studied exactly in \cite{Cremonini:2025yqe,Kanargias:2025vul}. Instead, it provides three complementary pieces of information: a direct four-dimensional derivation with the ultraviolet AdS$_4$ source kept explicit, an independent check of the $T_q/T$ asymptotics in the overlap regime, and a framework that can be generalized to geometries without a useful two-dimensional reduction.

The weak-coupling result is also naturally compared with the exact answers in terms of the energy above extremality. At fixed charge, the semiclassical equation of state gives $E\propto T^2$, and hence \eqref{eq:eta one loop} implies $\eta_{\rm one\mbox{-}loop}\propto E^{-1/2}$. In the strongly coupled Schwarzian regime, the exact equation of state becomes linear in $T$ and the exact calculations find the corresponding $T^{-1/2}$ behavior. Thus both regimes are compatible with an $E^{-1/2}$ dependence when their respective equations of state are used. This observation is a consistency check, not an extension of the one-loop validity domain.

\begin{figure}[H]
    \centering
    \includegraphics[width=0.82\linewidth]{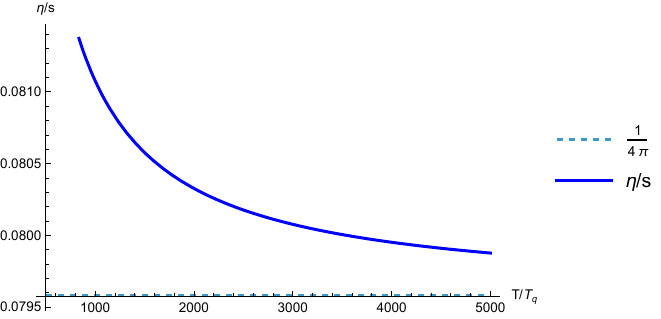}
    \caption{A controlled canonical-ensemble example of the one-loop corrected ratio. We take $r_0=L=1$, $\kappa_4=5\times10^{-4}$, and $V=3$, for which $T_q=2.5\times10^{-7}$. The plotted interval $500\leq T/T_q\leq5000$ also satisfies $1.25\times10^{-4}\leq TL^2/r_0\leq1.25\times10^{-3}$.}
    \label{fig:etaOvers}
\end{figure}

Figure \ref{fig:etaOvers} illustrates Eq.~\eqref{Eq:eta-over-s} using parameters that satisfy \eqref{eq:validity hierarchy} throughout the plotted interval. The correction is perturbative and approaches $1/(4\pi)$ as $T/T_q$ increases.

\paragraph{Comment on downward features of the curve} As mentioned above, a downward shift requires an exponentially large temperature, which may exceed the regime of validity of our calculation. However, at the risk of being imprecise, we comment that, by allowing the temperature to go above the second bound in \eqref{eq:validity hierarchy}, we can see some interesting features of the curve $\eta/s$. As shown in Figure \ref{fig:etaOvers,abuse}, the competition between the $1/T$ term and the $\log T$ term gives rise to a minimum in the curve, and $\eta/s$ approaches the classical $1/4\pi$ line from below.

\begin{figure}[H]
\centering
\includegraphics[width=0.82\linewidth]{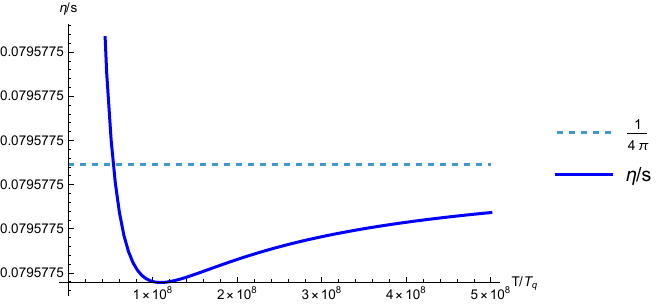}
\caption{The one-loop corrected ratio at high temperature. We take $r_0=L=1$, $\kappa_4=5\times10^{-4}$, and $V=3$, for which $T_q=2.5\times10^{-7}$. The plotted interval $5000\leq T/T_q\leq5\times10^8$ is equivalent to $1.25\times10^{-3}\leq TL^2/r_0\leq125$, which already exceeds the regime of validity in \eqref{eq:validity hierarchy}.}
\label{fig:etaOvers,abuse}
\end{figure}

\begin{figure}[H]
    \centering
    \includegraphics[width=0.82\linewidth]{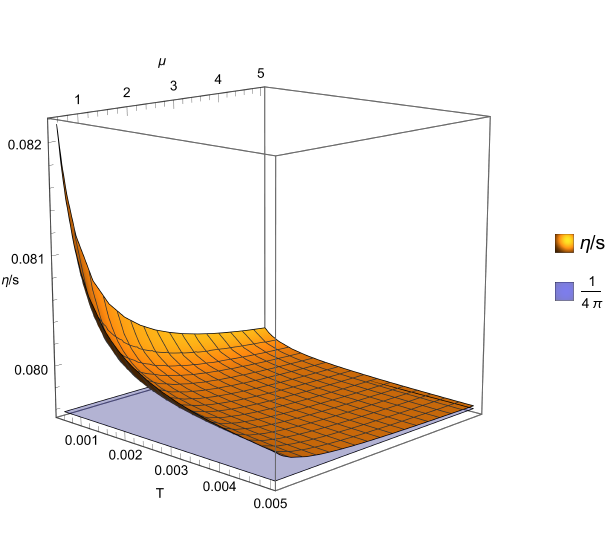}
    \caption{Grand-canonical illustration of the one-loop corrected ratio. Each constant-$\mu$ slice is evaluated at fixed chemical potential using $r_0=L^2\mu/\sqrt{3}$. We take $L=1$, $\kappa_4=5\times10^{-4}$, $V=3$, $0.5\leq\mu\leq5$, and $5\times10^{-4}\leq T\leq5\times10^{-3}$. The whole displayed domain satisfies $T/T_q^{(\mu)}\gtrsim1.15\times10^3$, and $TL^2/r_0\lesssim1.8\times10^{-2}$. The translucent plane denotes $1/(4\pi)$.}
    \label{fig:etaOvers2D}
\end{figure}

\paragraph{Chemical potential and grand canonical ensemble}
The analytic derivation above uses the canonical ensemble. To display the role of the chemical potential, Figure \ref{fig:etaOvers2D} is instead organized grand canonically: each slice of constant $\mu$ is evaluated at fixed chemical potential. In this ensemble,
\begin{equation}
    r_0=\frac{L^2\mu}{\sqrt{3}},\qquad
    r_+=r_0+\frac{2\pi L^2}{3}T+\mathcal{O}(T^2).
\end{equation}
To get the plot for grand canonical ensemble, instead of repeating the above calculation once again, we take the following short cut:

In the above calculation of the lifted eigenvalues of the Schwarzian modes \eqref{Eq: whole lifted eigenvalue}, the second piece $\delta\Lambda_n[\delta g]$ is obtained at extremality, which is independent of the ensemble choice. Therefore, the second piece of \eqref{Eq: whole lifted eigenvalue} should be the same in a grand canonical ensemble. For the first piece $\delta \Lambda_n[g^{(1)}]$, we calculate it in terms of $r_0$ and $T$. In the canonical ensemble we set $q=q_0=\sqrt{3} r_0^2$, while the chemical potential is a function of temperature, $\mu=\mu(r_0,T)$. To change to a grand canonical ensemble, one can still use $r_0$ and $T$ as parameters while setting $\mu=\mu_0=\tfrac{\sqrt{3}r_0}{L^2}$ and $q=q(r_0,T)$. The first order metric perturbation $g^{(1)}$ should also be changed to the grand canonical one. In doing this, one will find $\delta \Lambda_n[g^{(1)}]$ takes exactly the same form. This is actually a generic feature that $\delta \Lambda_n[g^{(1)}]$ is ensemble independent. Reasons behind this are discussed in \cite{PandoZayas:2026vbg}. Therefore, to get a grand canonical result of $\eta/s$, we can just take \eqref{Eq:eta-over-s} and keep the same expression for the one-loop part of $\eta$. The common tree-level factor in $\eta_{\rm tree}$ and $s_{\rm tree}$ becomes
\begin{equation}
     A_{\mu}(T)=1+\frac{4\pi L^2T}{3r_0}+\mathcal{O}(T^2).
 \end{equation}
The scale in the logarithm in \eqref{Eq: tq}, by the inverse linear entropy coefficient in the chosen ensemble, is then
\begin{equation}
    T_q^{(\mu)}=\frac{3\kappa_4^2}{2r_0V}
    =\frac{3\sqrt{3}\,\kappa_4^2}{2L^2V\mu}.
\end{equation}
The factor of two relative to the canonical value is part of the ensemble-dependent thermodynamic normalization; it changes only an additive $\mathcal{O}(1)$ term in the logarithm. The lifted eigenvalue and the explicit corrections $\eta_{\rm one\mbox{-}loop}=9\pi r_0/(L^2TV)$ and $s_{\rm one\mbox{-}loop}=3\log(T/T_q^{(\mu)})/(2V)$ retain the same form. To be consistent, we also replace the $T_q$ in the logarithm by $T_q^{(\mu)}$. Then the ratio becomes
\begin{equation}\label{Eq:eta-over-s,grand canonical}
\begin{aligned}
    \frac{\eta}{s}=& \frac{1}{4\pi}\,\,\times \,\,
\frac{1+\frac{9\pi r_0}{L^2TV}\big[\frac{1}{2\kappa_4^2}(\frac{r_0}{L})^2\left(1+\frac{4\pi L^2T}{3r_0}\right)\big]^{-1}}{1+\frac{3}{8\pi V}\log{\frac{T}{T_q^{(\mu)}}}\big[\frac{1}{2\kappa_4^2}(\frac{r_0}{L})^2\left(1+\frac{4\pi L^2T}{3r_0}\right)\big]^{-1}}.
\end{aligned}
\end{equation}
The controlled grand-canonical window is
\begin{equation}
\label{eq:grand canonical window}
    \frac{3\sqrt{3}\,\kappa_4^2}{2L^2V\mu}\ll T\ll\frac{\mu}{\sqrt{3}}.
\end{equation}
Increasing $\mu$ widens this window and suppresses the relative one-loop correction, so that the ratio approaches its classical value. This plot is intended to illustrate the parameter dependence of the Kubo-defined zero-frequency response. We do not infer from it that conventional hydrodynamics remains valid outside the hierarchy appropriate to a large system and sufficiently large chemical potential \cite{Davison:2013bxa,Nian:2025oei}.

\paragraph{Thermodynamic limit.}
The modes retained in this calculation are independent of $x$ and $y$ and constitute a single homogeneous collective sector. Their contribution scales as
\begin{equation}
\label{eq:volume scaling}
    T_q\sim\frac{1}{V},\qquad
    \eta_{\rm one\mbox{-}loop}\sim\frac{1}{TV},\qquad
    s_{\rm one\mbox{-}loop}\sim\frac{1}{V}\log\left(\frac{T}{T_q}\right).
\end{equation}
Consequently,
\begin{equation}
    \lim_{V\to\infty}\frac{\eta}{s}=\frac{1}{4\pi}
\end{equation}
at fixed nonzero temperature and fixed background densities. The result derived here is therefore a finite-volume correction to the zero-momentum response. A distinct near-extremal double-scaling limit can be defined by taking $V\to\infty$ and $T\to0$ with $T/T_q$ fixed and large; in that limit the relative correction to $\eta$ remains finite. This statement concerns only the infrared-enhanced homogeneous Schwarzian contribution and does not exclude ordinary volume-independent bulk-loop corrections from other modes.

\section{Conclusions}

We have computed the infrared-enhanced one-loop contribution of the homogeneous Schwarzian sector to a stress-tensor correlation function of a near-extremal asymptotically AdS$_4$ black brane. The calculation is performed directly in four dimensions with a finite transverse torus and with the source normalized at the ultraviolet AdS$_4$ boundary. At tree level, the small-temperature expansion reproduces $\eta/s=1/(4\pi)$. At one loop, the shear source couples only to the $n=2$ lifted zero mode and produces the correction $\eta_{\rm one\mbox{-}loop}\propto T_q/T$.

The result is perturbative in the hierarchy $T_q\ll T\ll r_0/L^2$. It agrees with the weak-coupling asymptotics of the exact two-dimensional treatments while providing an independent four-dimensional derivation. Because the relevant mode is spatially homogeneous, the correction is subextensive and vanishes in the strict $V\to\infty$ limit at fixed temperature. We have therefore interpreted the result as a finite-volume correction to the Kubo-defined zero-frequency response, rather than as a generic or parametrically sizable low-temperature violation of the KSS lower bound.

The semiclassical universality of the ratio is related to the connection between shear viscosity and the low-frequency absorption cross section \cite{Son:2002sd,Das:1996we}. The quantum corrections localized in the near-extremal throat suggest a corresponding relation to the quantum absorption problem \cite{Emparan:2025sao,Biggs:2025nzs,Emparan:2025qqf,Betzios:2025sct}. It would be valuable to study this relation directly in four dimensions and to extend the source-dependent determinant analysis to additional channels and to rotating backgrounds. Connection-coefficient methods may provide a useful framework for such extensions \cite{Aminov:2020yma,Aminov:2023jve}.

\subsection*{Note added}

After the first version of this paper appeared on arXiv, Refs.~\cite{Cremonini:2025yqe,Gouteraux:2025exs,Kanargias:2025vul} studied related observables using dimensionally reduced near-AdS$_2$ descriptions and exact throat correlators. These methods access the strongly coupled regime $T\ll T_q$, which lies outside the one-loop domain of the present calculation. In the overlapping weak-coupling regime $T\gg T_q$, Refs.~\cite{Cremonini:2025yqe,Kanargias:2025vul} find the same $T_q/T$ scaling. In the strongly coupled regime they find a $\sqrt{T_q/T}$ behavior, in agreement with the quantum absorption analysis of \cite{Emparan:2025sao}. Ref.~\cite{Gouteraux:2025exs} obtains a different scaling associated with a different choice of the infrared conformal dimension.

At fixed charge, the weak-coupling equation of state gives $E\propto T^2$, so our result can be written as $\eta_{\rm one\mbox{-}loop}\propto E^{-1/2}$. The exact strong-coupling result has the same dependence on energy after the quantum-corrected equation of state is used. We regard this as a useful consistency check across the overlapping descriptions, while maintaining that the direct one-loop derivation itself is valid only for $T\gg T_q$.

\section*{Acknowledgments}
We are thankful to Giulio Bonelli, Sera Cremonini,  Blaise Gout\'eraux,  Xiao-Long Liu, Sabyasachi Maulik, Jun Nian, Koenraad Schalm, Alessandro Tanzini, Yu Tian, Joaqu\'\i n Turiaci, Congyuan Yue and Hongbao Zhang for clarifying discussions or ongoing explorations in related topics. 
This work is partially supported by the U.S. Department of Energy under grant DE-SC0007859. J.Z. acknowledges support from the Leinweber Institute for Theoretical Physics in the form of Summer Fellowships.

\appendix
\section{The structure of the Lichnerowicz operator}
\label{appendix:Lich operator}

The generalized Lichnerowicz operator is obtained by the second-order metric perturbation of the action. In this appendix, we list the ingredients of the operator used in the previous sections.

\begin{align}
\nonumber
   h_{\alpha\beta}\Delta^{\alpha \beta, \mu \nu}_{EH}h_{\mu\nu}=& h_{\alpha \beta}\Big(\frac{1}{2} g^{\alpha \mu} g^{\beta \nu} \square - \frac{1}{4} g^{\alpha \beta} g^{\mu \nu} \square + R^{\alpha \mu \beta \nu}\Big)\\
&+ R^{\alpha \mu} g^{\beta \nu}-R^{\alpha \beta} g^{\mu \nu}-\frac{1}{2} R g^{\alpha \mu} g^{\beta \nu}+\frac{1}{4} R g^{\alpha \beta} g^{\mu \nu}) h_{\mu \nu},\\
h_{\alpha \beta} \Delta_{1}^{\alpha \beta, \mu \nu} h_{\mu \nu}=&h_{\alpha\beta}(-\frac{1}{2}g^{\alpha\mu}g^{\beta\nu}+\frac{1}{4}g^{\alpha\beta}g^{\mu\nu})h_{\mu\nu},
\\
\nonumber
h_{\alpha \beta} \Delta_{F}^{\alpha \beta, \mu \nu} h_{\mu \nu}=&h_{\alpha \beta}\,\Bigl( -\frac{1}{8} F_{\rho\sigma}F^{\rho\sigma}\left(2 g^{\alpha \mu} g^{\beta \nu}-g^{\alpha \beta} g^{\mu \nu}\right)+ F^{\alpha \mu} F^{\beta \nu}\\
&+2 F^{\alpha \gamma} F^{\mu}{}_{\gamma} g^{\beta \nu}-F^{\alpha \gamma} F^{\beta}{}_{\gamma} g^{\mu \nu} \Bigr) h_{\mu \nu},\\
\nonumber
\end{align}
For the readers' convenience, here we record the intermediate result in calculating \eqref{Eq: Lifted eigenvalue by T}:
\begin{equation}
\begin{aligned}
        &\int dx^4 \sqrt{g^{(0)}}h^{(n)*}_{\alpha\beta}\delta\Delta_L^{\alpha\beta,\mu\nu} h^{(n)}_{\mu\nu}\\&=\int_1^\infty dY
\frac{16\pi T \, l^{2}  n  \left(-1+n^{2}\right) 
\left(\frac{-1+Y}{1+Y}\right)^{n}
\left(4 n^{2}(2+Y)+2(-1+Y)^{2}(4+Y) 
- 5n(-3+2Y+Y^{2})\right)}
{r_{0}^{3}  (-1+Y)^{2}  (1+Y)^{4}}
\end{aligned}
\end{equation}
It is quite remarkable that after integration the result is proportional to $n$ despite the quadratic dependence in the integrand; this leads to a simple evaluation using zeta-function regularization. In deriving the above result, one can make use of the structures uncovered in \cite{Maulik:2024dwq}, according to which only two terms in the Lichnerowicz operator contribute to the shift in eigenvalues. 
\section{Rotation back to Lorentzian time}
\label{appendix: Integral}
In this appendix, we provide details of the calculation leading to \eqref{Eq: Lifted Eigenvalue by fluctuation}. Our goal is to evaluate the integral \eqref{Eq: integral} on the background of $g=g^{(0)}+\delta g$. Similar to the  calculation of the one-loop correction to the entropy in Section \ref{Sec: One-loop partition function}, we do the integral only in the near-horizon ``throat", as this is where the quantum effect enters. 
Substituting \eqref{Eq: the AdS/CFT modes}, direct algebra shows 
\begin{equation}
\begin{aligned}
h^{(n)*}_{\alpha\beta}\delta\Delta_L^{\alpha\beta,\mu\nu} h^{(n)}_{\mu\nu}\;=\; 
& -\,\frac{9\, i\, l^{2}\, n \,(n^{2}-1)(n - 2Y)}%
{8 \, L^{2} \pi^{8} r_{0}^{2} T V \,(1-Y)^{3}(1+Y)^{3}} 
\left(\frac{1-Y}{1+Y}\right)^{n}
\\[6pt]
& \times 
\Bigg[
\int d\omega_{1}\, d k_{x1}\, d k_{y1}\;
e^{-i k_{x1} x - i k_{y1} y - \tfrac{\tau \omega_{1}}{2\pi T}}
\left(\frac{1-Y}{1+Y}\right)^{\tfrac{i \omega_{1}}{4\pi T}}
\, \phi(\omega_{1},k_{x1},k_{y1})
\Bigg]
\\[6pt]
& \times 
\Bigg[
\int d\omega_{2}\, d k_{x2}\, d k_{y2}\;
e^{-i k_{x2} x - i k_{y2} y - \tfrac{\tau \omega_{2}}{2\pi T}}
\left(\frac{1-Y}{1+Y}\right)^{\tfrac{i \omega_{2}}{4\pi T}}
\, \omega_{2}\, \phi(\omega_{2},k_{x2},k_{y2})
\Bigg],
\end{aligned}
\end{equation}
where we keep only the leading-order contribution. Unlike the calculation of the lifted eigenvalue \eqref{Eq: Lifted eigenvalue by T}, which starts at linear order, here the non-zero contribution starts at quadratic order.

In the Euclidean near-horizon throat the integral measure is 
\begin{equation}
    \int d^4x \sqrt{g^{(0)}}=\int_0^{2\pi}d\tau\int_1^{\infty}dY\int dxdy\sqrt{g^{(0)}}
\end{equation}
However, in order to get the correct momentum space Green's function, we need to rotate back to the Lorentzian time $t=\tfrac{-i\tau}{2\pi T}$. The above integral measure becomes
\begin{equation}
    \int d^4x \sqrt{g^{(0)}}\to\int_1^\infty dY\int dtdxdy \sqrt{-g_L^{(0)}}=\int_1^\infty dY\int dtdxdy \;2\pi T\sqrt{g^{(0)}},
\end{equation}
where $g_L^{(0)}$ is the metric determinant after we rotate back to Lorentz signature
\begin{equation}
    g^{(0)}_{\tau\tau}d\tau^2\to-g^{(0)}_{\tau\tau}(2\pi T)^2dt^2.
\end{equation}
Then we pick out the two exponentials and integrate over the boundary dimensions ${t,x,y}$,
\begin{equation}
    \int dtdxdy\; e^{ -i \omega_{1}t+i k_{x1} x +i k_{y1} y}e^{ -i \omega_{2}t+i k_{x2} x +i k_{y2} y}=(2\pi)^3\delta(\omega_1+\omega_2,k_{x1}+k_{x2},k_{y1}+k_{y2})
\end{equation}
The delta function eliminates one integral in the three-dimensional momentum space and constrains $\omega_1=-\omega_2,k_{x1}=-k_{x2},k_{y1}=-k_{y2}$.
Using $\sqrt{g^{(0)}}=\tfrac{r_0^2}{6}$, the lifted eigenvalues can be shown as 
\begin{equation}
\begin{aligned}
        \delta\Lambda_n[\delta g]=-\int_1^\infty dY\frac{3 i\, l^{2} n (n^{2}-1)(n-2Y)}{L^{2}\pi^{4}\, V \,(1-Y)^{3}(1+Y)^{3}}
\left(\frac{1-Y}{1+Y}\right)^{\!n}
\\ &\quad\times\int d\omega d k_{x} d k_{y}\;
\omega\,\phi(-\omega,-k_{x},-k_{y})\,\phi(\omega,k_{x},k_{y})
\end{aligned}
\end{equation}
Performing the integral over $Y$ gives \eqref{Eq: Lifted Eigenvalue by fluctuation}.
\bibliographystyle{JHEP}
\bibliography{Hawk-Kerr}
\end{document}